\documentclass[reprint,twocolumn,prb]{revtex4-1}
\usepackage{siunitx}
\usepackage{miller}
\usepackage{graphicx}
\usepackage{hyperref}

\begin{document}
\title{Vibrations of single-crystal gold nanorods and nanowires}

\author{L. Saviot}
\email{lucien.saviot@u-bourgogne.fr}
\affiliation{
Laboratoire Interdisciplinaire Carnot de Bourgogne,
UMR 6303 CNRS--Universit\'e Bourgogne Franche-Comt\'e,
9 Avenue Alain Savary,
BP 47 870,
21078 Dijon Cedex,
France
}

\begin{abstract}
The vibrations of gold nanowires and nanorods are investigated numerically
in the framework of continuum elasticity using the Rayleigh-Ritz variational method.
Special attention is paid to identify the vibrations relevant in Raman scattering experiments.
A comprehensive description of the vibrations of nanorods is proposed
by determining their symmetry, comparing with standing waves in the corresponding nanowires
and estimating their Raman intensity.
The role of experimentally relevant parameters such as the anisotropic
cubic lattice structure, the presence of faceted lateral surfaces
and the shape of the ends of the nanorods is evaluated.
Elastic anisotropy is shown to play a significant role contrarily to the presence of facets.
Localized vibrations are found for nanorods with flat ends.
Their evolution as the shape of the ends is changed to half-spheres is discussed.
\end{abstract}

\maketitle

\section{Introduction}

Non-spherical gold nanoparticles have attracted considerable attention during the last few decades
mainly because their optical properties are strongly shape-dependent.
Gold nanorods (NRs) in particular have been the focus of many studies,
some dedicated to understanding and controlling their synthesis\cite{LohseCM2013}
while other focused on their properties.
In this context, their acoustic vibrations have been investigated as a mean to measure
the dimensions of the NRs, but also to study their acousto-plasmonic coupling with the
localized surface plasmon resonance (LSPR).\cite{LargeNL09,CrutPR2015}
Transient absorption measurements are a tool of choice in this context in particular because
single particle measurements are possible.\cite{HuJACS2003}
It enables to observe totally symmetric vibrations, \textit{i.e.}, the extensional
vibration which consists in an oscillation of the length of the NRs
and the breathing vibration which is an oscillation of the radius.
Low-frequency Raman scattering is also of interest as it obeys distinct selection rules.
The same totally symmetric vibrations and others non-totally symmetric ones have been observed experimentally.\cite{MargueritatAPA07,MargueritatNL06,LargeNL09}
A continuous-four-wave mixing approach has also been proposed recently.\cite{WuOE2016}
It was shown to be sensitive to some Raman active vibrations.
Finally, another experimental technique known as extraordinary acoustic Raman spectroscopy was also proposed.\cite{WheatonNP2015}
It is a promising approach enabling single particle measurements
which has not been applied to gold NRs yet.

Experimental measurements must be supported by calculations to understand the nature of the observed vibrations.
Most often, continuum isotropic elasticity is assumed for NRs made of gold and other materials
resulting in models based on analytic expressions in cylindrical
coordinates.\cite{BalandinJBN05}
While this assumption is sometimes valid, it is also known to be a poor one in many cases.
This is expected in particular for ultrathin gold nanowires (NW)\cite{PazosPerezL2008} and NRs
which are single-crystalline.
Gold is strongly anisotropic (Zener ratio: 2.9)
which results for example in a large splitting of the quadrupolar
vibrations of nanospheres (NS).\cite{PortalesPNAS08,SaviotPRB09}
Such a splitting must also exist for gold NRs but calculations of this splitting are lacking.
In addition, anisotropic elasticity has also an impact on the other vibrations.
It shifts the frequency of the extensional vibration significantly enough that
taking into account elastic anisotropy is required.\cite{GoupalovNL14,GoupalovNL14cor}
Anisotropy also results in the existence of several vibrations sharing the breathing character
and therefore a broadening of the corresponding peak.
This effect must be properly taken into account before discussing damping.\cite{YuNL2013}
Finally, the impact of non-circular cross-sections as observed in penta-twinned gold NRs\cite{GanPCCP16}
or single-cristalline NRs must also be evaluated.

The purpose of this work is to investigate numerically the vibrations of gold NRs
and in particular those which play a role in experiments.
Vibrations of gold NWs are also calculated.
The influence of elastic anisotropy, of the cross-section of the NWs
and of the ends of the NRs are investigated.
The frequencies of the NRs are also compared with those of standing waves built from propagating waves
of the NWs in order to provide a simple scaling law.
A similar approach has been used before to establish the effective wavelength scaling for optical nanoantennas made of gold NRs.\cite{NovotnyPRL07,BryantNL08}

\section{Methods}
In the following, the radius for the NSs and the circular NRs and NWs is
kept constant at $R=\SI{5}{nm}$ and their main axis is along \hkl[001].
Octagonal NWs are also considered.
Their surface is delimited by \hkl{730} planes as observed in elongated tetrahexahedral gold nanocrystals.\cite{RajendraNS16}
The half-lengths along the \hkl<100> and \hkl<110> directions
($d\textsubscript{100}/2=\SI{5.296}{nm}$ and $d\textsubscript{110}/2=\SI{5.243}{nm}$)
were chosen so that the surface area of the cross-section of all the NWs is the same.
The results obtained for the dimensions given above can be used to determine the frequencies for different dimensions.
The frequencies of the NSs scale as $1/R$.
The same scaling applies to NRs provided the aspect ratio (length/diameter) is preserved.
For NWs, both the frequencies and the wavevectors scale as $1/R$.
The stiffness tensor for gold is defined by $C_{11}=\SI{191}{GPa}$, $C_{12}=\SI{162}{GPa}$ and $C_{44}=\SI{42.4}{GPa}$, and its mass density is $\rho=\SI{19.283}{g/cm^3}$.\cite{LBAu}
For the isotropic approximation of gold, the stiffness tensor
obtained from the 3D averaged sound speeds is given by
$C_{11}=\SI{213.3}{GPa}$ and $C_{12}=\SI{152.6}{GPa}$.

Calculations for NWs are performed by taking into account the translational symmetry
starting from the method proposed by \citet{NishiguchiJPCM97} for rectangular quantum wires.
The calculations for circular NWs with isotropic elasticity are also
checked with exact analytical expressions.\cite{BalandinJBN05}
Due to the translational symmetry, the vibrations of NWs are not discrete and form branches
for varying wavevectors $q$.
The vibrations of the NRs and NSs are obtained using the Rayleigh-Ritz variational method\cite{VisscherJASA1991}
to take into account anisotropy due to the non-spherical shape or cubic elasticity.
Similar calculations exist in the literature mostly for isotropic NRs.\cite{ZemanekJASA72,GrinchenkoSAM1980,LeissaJASA1995,HeyligerJASA2003,JaglinskiRSI2011}
The present work improves on such previous works in several ways.
All the vibrations are considered up to the first breathing-like mode.
Totally symmetric and non-totally symmetric vibrations are investigated.
The irreducible representations of the vibrations of the NRs and NSs are determined
in order to distinguish them and also to bring out the vibrations of interest
such as the Raman active ones.\cite{SaviotPRB09}
Similarly, the irreducible representation of the $q=0$ phonons of the NWs are determined
for the same point group as the one of the corresponding NRs,
namely D\textsubscript{$\infty$h} for isotropic elasticity and circular cross-sections and D\textsubscript{4h} otherwise,
by taking into account a single symmetry plane perpendicular to the symmetry axis at $z=0$.
In addition, the variation of the volume of the NSs and NRs during the vibration is calculated
to help identify breathing vibrations.\cite{SaviotPRB09}
For comparison, the variation of the surface area of the cross-section is also calculated for NWs.
Calculations for NWs are then used to describe the vibrations of NRs in terms
of confined vibrations (standing waves) when possible.

\section{Results and Discussion}

\subsection{Nanowires}

\begin{figure*}
\includegraphics[width=0.75\textwidth]{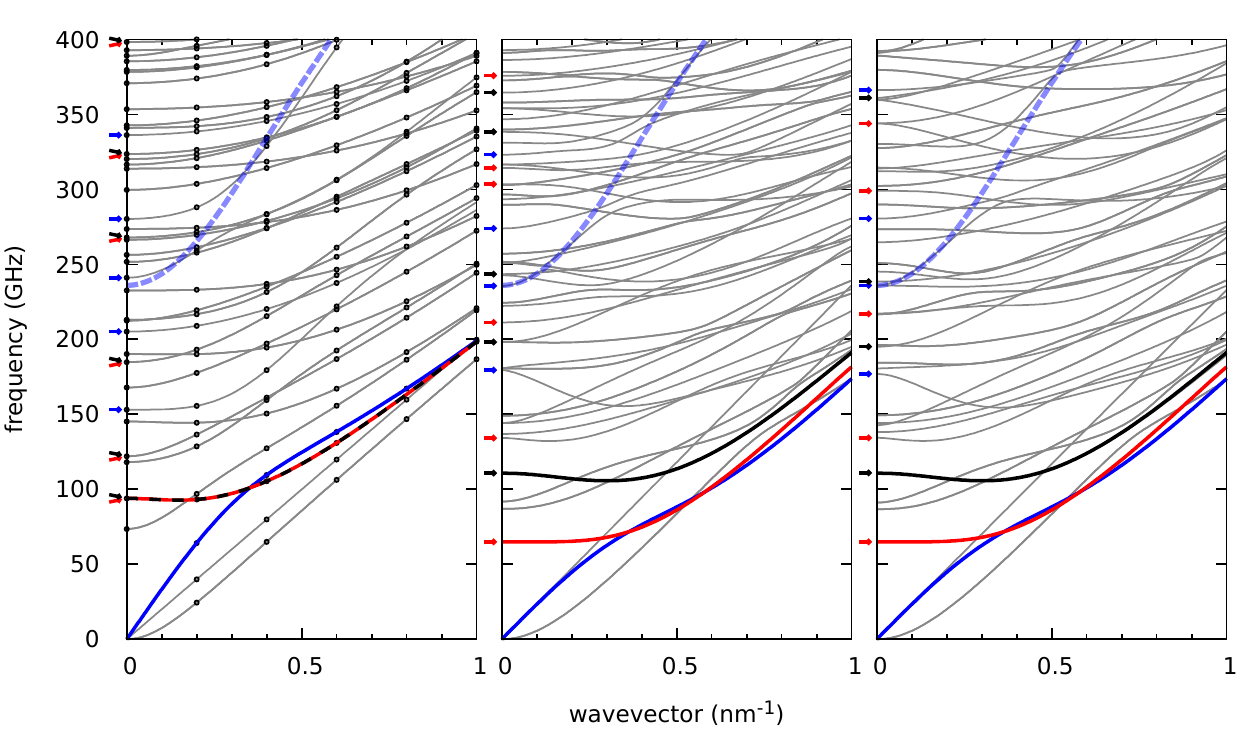}
\includegraphics[width=0.5\textwidth]{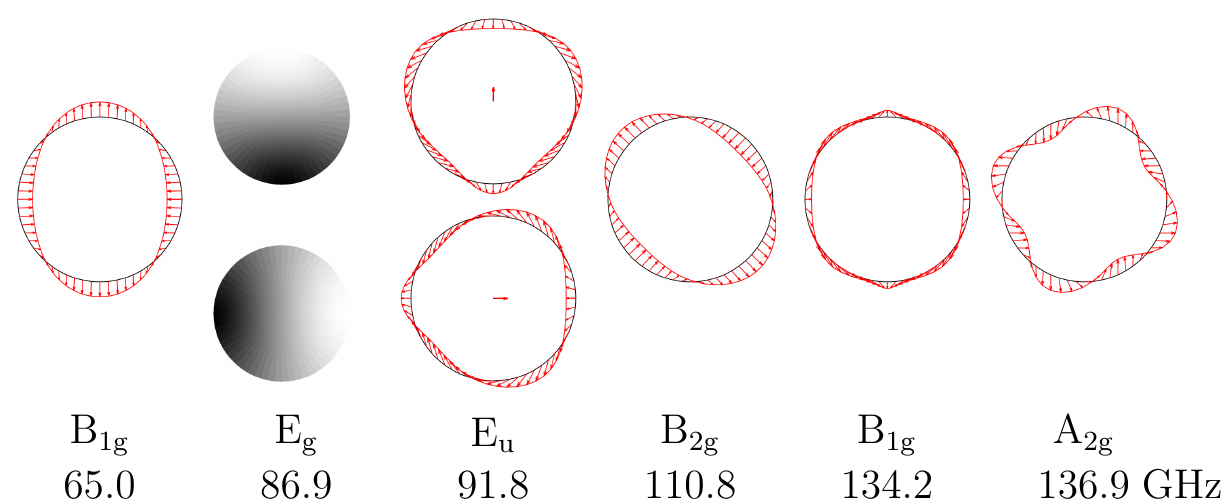}
\caption{Top: phonon dispersion for the isotropic circular NW (left)
and the anisotropic circular (center) and octagonal (right) NWs.
Analytical calculations for the isotropic circular NW are shown with circles (left)
for a few selected $q$.
Zone center A\textsubscript{1g} modes are shown with blue arrows,
B\textsubscript{1g} with red arrows and B\textsubscript{2g} with black arrows.
The dilatational branch is colored in blue, the branch starting from the lowest frequency B\textsubscript{1g} mode is colored in red and the B\textsubscript{2g} one in black.
The isotropic E\textsubscript{2g} modes are shown with both red and black arrows
and the branch starting from the lowest frequency E\textsubscript{2g} mode is dashed
in red and black (left).
The same dashed thick blue curve is plotted as a guide for the eye for a breathing branch
without anti-crossing patterns.\\
Bottom: displacement in the $xy$ plane of the lowest frequency $q=0$ phonons of the anisotropic circular cylinder.
For the E\textsubscript{g} vibrations, the displacements are along $z$ and the
black and white areas move in opposite directions.
}
\label{disp}
\end{figure*}

Fig.~\ref{disp} presents the phonon branches for the circular NW with isotropic gold and the circular and octagonal NWs with cubic gold.
For the Rayleigh-Ritz variational calculation, the displacements were expanded on the $x^l y^m e^{iqz}$
basis with $l$ and $m$ being integers and $0 \leq l+m \leq 23$.
For the octagonal NW, numerical issues with the Cholesky decomposition in the LAPACK ZHEGV routine\cite{laug}
restricted this range to $0 \leq l+m \leq 9$.
We first note the excellent agreement for the isotropic circular NW with the exact analytic
calculations.\cite{BalandinJBN05}
Only a few selected $q$ are shown in Fig.~\ref{disp} (left) for the analytic method but the agreement
is excellent over the 0--\SI{1}{nm^{-1}} range ($\Delta\nu/\nu<10^{-7}$).

\begin{table}
\begin{tabular}{rr|r|r|r}
 & \multicolumn{2}{c|}{breathing} & \multicolumn{2}{c}{quadrupolar}\\
 & isotropic & cubic & isotropic & cubic\\
\hline
NS  & 310.93   &  310.12             &  106.20  & E\textsubscript{g} \hspace{0pt plus 1filll}  74.62\\
    & $\ell=0$ & A\textsubscript{1g} & $\ell=2$ & T\textsubscript{2g} \hspace{0pt plus 1filll} 120.47\\
\hline
circular NW & 241.06 &  235.63             & 93.94 & B\textsubscript{1g} \hspace{0pt plus 1filll} 64.99\\
$q=0$    & $m=0$  & A\textsubscript{1g} & $m=2$ & B\textsubscript{2g} \hspace{0pt plus 1filll} 110.77\\
    \hline
octagonal NW & 240.50              & 235.97              & B\textsubscript{1g} \hspace{0pt plus 1filll} 93.80 & B\textsubscript{1g} \hspace{0pt plus 1filll} 64.93 \\
$q=0$    & A\textsubscript{1g} & A\textsubscript{1g} & B\textsubscript{2g} \hspace{0pt plus 1filll} 94.11 & B\textsubscript{2g} \hspace{0pt plus 1filll} 110.88\\
\end{tabular}

\caption{Breathing(-like) and quadrupolar(-like) frequencies for NSs, circular and octagonal NWs
with isotropic and cubic elasticity.
The irreducible representations are given (D\textsubscript{4h} point group),
$\ell$ is given for the spherical symmetry (isotropic NS)
and $m$ for the cylindrical one (isotropic circular NW).
\label{brqu}
}
\end{table}

The breathing and quadrupolar-like frequencies for NSs and NWs
with isotropic and cubic elasticity are given in Table~\ref{brqu}.
For the isotropic NS, only the spheroidal $\ell=0$ (breathing) and
$\ell=2$ (quadrupolar) vibrations are Raman active.
In the other cases, the modes were identified by checking the associated displacements.
When changing the shape from a NS to a NW, the breathing frequency decreases much more
(\SI{-70}{GHz}) than the quadrupolar-like frequency (\SI{-10}{GHz}).

Introducing elastic anisotropy
\footnote{The slope at $q=0$ of the torsional and longitudinal branches
are identical for anisotropic NWs in Fig.~\ref{disp}. This is an accidental coincidence
for the $C_{ij}$ chosen in this work.}
results in a significant splitting of
the quadrupolar-like modes as already reported for NSs.\cite{SaviotPRB09}
It confirms that elastic anisotropy must be taken into account for NWs too.
As was discussed in a previous work for NS\cite{StephanidisPRB07}, this splitting
can be understood as transverse waves with sound speeds
$\sqrt{\frac{C_{11}-C_{12}}{2\rho}}$
and $\sqrt{\frac{C_{44}}{\rho}}$ confined over the same distance (diameter).
The ratio of these sound speeds is the square root of the Zener anisotropy ratio
(1.7105 for cubic gold).
In this simple approach, it is equal to the frequency ratio after
splitting.
Indeed, the ratio of the $q=0$ B\textsubscript{2g} and B\textsubscript{1g} frequencies
is 1.7044.
Therefore, the square of this ratio as measured for example from Raman
spectra is expected to be an accurate estimate of the Zener anisotropy
ratio.

The difference between the octagonal and circular NWs is very small.
Again, the same conclusion was reached for nanocrystals having similar shapes
(sphere, cuboctahedron and truncated cuboctahedron) provided the volume was the same.\cite{SaviotPRB09}
For this reason, only circular NWs and NRs are considered in the following.
The phonon branches starting with the NWs vibrations listed in Table~\ref{brqu} are
highlighted in Fig.~\ref{disp}.
The branch starting from the $q=0$ breathing vibration is not highlighted in the same way
because it couples with other branches making its dispersion interesting
up to the first anti-crossing feature only.
Instead, a dashed thick blue curve is plotted as a guide for the eye.
It was obtained by following the modes having the largest surface area variation
during vibration.
It was determined for the anisotropic circular NW and fitted to
$\omega_\text{breathing} = 236  + 782 q^2 -1207 q^4 + 945 q^6$ where $\omega$ is in GHz and $q$ in nm\textsuperscript{-1}.
The same curve is plotted for the three NWs.
The breathing modes are very similar for the three NWs except
for the position of the anti-crossing patterns.

\subsection{Nanorods}
The calculation for the NRs were performed by expanding the displacements on the $x^l y^m z^n$
basis,\cite{VisscherJASA1991} where $l$, $m$ and $n$ are integers and $0 \leq l+m+n \leq 20$.
The NRs have a circular cross-section and are made of gold with cubic elasticity.
Therefore, the vibrations will be compared to those of the circular NW with cubic elasticity.
Calculations were performed for NRs with straight ends (NW cut by two planes perpendicular
to the \hkl[001] symmetry axis) and also for NRs with half-sphere ends.
In the following, $L$ always corresponds to the total length of the NRs.

\subsubsection{Extensional modes}
\begin{figure}
\includegraphics[width=\columnwidth]{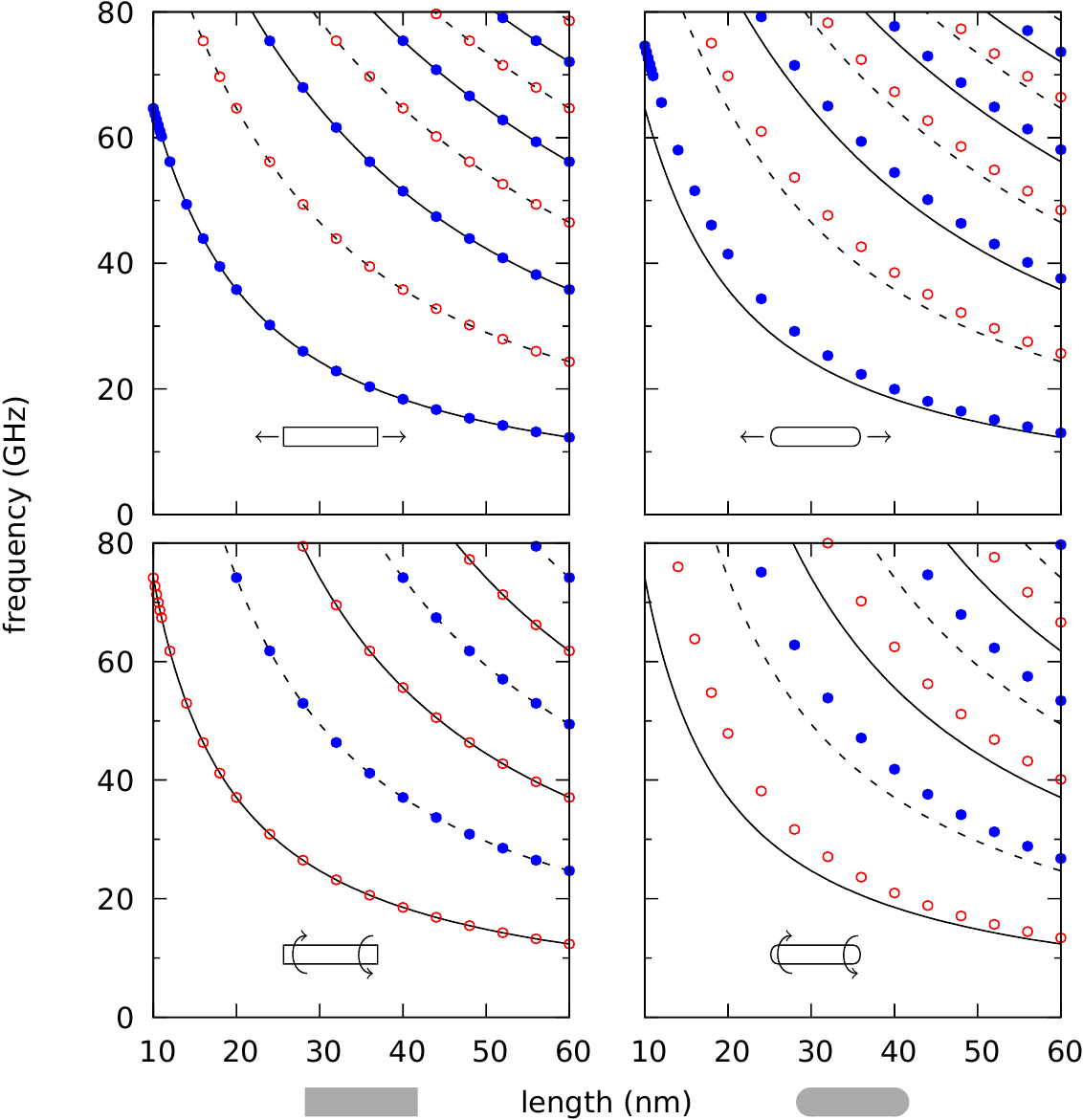}
\caption{Frequencies of the extensional (top) and torsional (bottom) vibrations
of anisotropic circular NRs as a function of their length $L$.
The NRs ends are flat (left) or half-spheres (right).
The lowest frequency even vibrations are plotted with full circles(blue online)
and the odd ones with empty circles (red online).
The frequencies derived from the dilatational phonon branch of the NW at $q=n\frac{\pi}{L}$
are plotted with continuous lines for odd $n$ and dashed lines for even $n$.
}
\label{ext}
\end{figure}

Let us start by considering the extensional modes of the NRs.
They correspond to dilatational phonons of the NWs (blue curves in Fig.~\ref{disp})
confined along the length of the NR.
Their frequencies can be derived assuming free boundary conditions at both ends.
For a NR with flat ends and considering first the totally symmetric modes which are Raman active,
$q=(2 n + 1)\frac{\pi}{L}$ with $n=0, 1, \ldots$
We compare the frequencies of the lowest A\textsubscript{1g} vibrations with this expression
in Fig.~\ref{ext} (top left).
The agreement is remarkable for the first few overtones.
A very good agreement is also obtained for the lowest frequency A\textsubscript{2u} vibrations
(Raman inactive) which are their anti-symmetric analogues corresponding to $q=2 n\frac{\pi}{L}$.

This first successful comparison demonstrates that determining the phonon
frequencies of the NW as a function of $q$ (in this case for the dilatational banch)
can be sufficient to calculate the frequency of some vibrations of the NRs with flat ends 
(in this case the extensional ones)
for arbitrary diameters and lengths.
The agreement is very good including above \SI{40}{GHz}
where the NW dilatational phonon frequency is no longer proportional to $q$.
This frequency range is relevant for short NRs and for overtones.

In Fig.~\ref{ext} (top right), the same approach is used for NRs with half-sphere ends.
The agreement is not as good as before, but the simple calculation using the NW dilatational
branch still provides a rather good approximation of the frequency.
The agreement is improved for large $L$ by using $q$ determined from the equivalent length
$L_{eq} = L - \frac{2}{3} R$ which corresponds
to the length of a NR with flat ends having the same volume.

\subsubsection{Torsional modes}
A similar treatment can be applied to torsional vibrations which consist in rotations
around the symmetry axis of the NRs.
These vibrations are of course associated to the NW torsion phonon branch
whose frequency is proportional to $q$ (linear variation in Fig.~\ref{disp}).
Odd torsional vibrations of the NRs correspond to the A\textsubscript{2g} irreducible representation
and even vibrations correspond to A\textsubscript{1u}.
The confinement along $L$ results in $q=n\frac{\pi}{L}$ with odd values of $n$ for A\textsubscript{1u}
and even values for A\textsubscript{2g}.
Fig.~\ref{ext} (bottom) shows the resulting frequencies for both kinds of NRs.
As before, the agreement for NRs with straight ends is remarkable.
However, the deviation observed for NRs with half-sphere ends is larger.

\subsubsection{Bending modes}
\begin{figure}
\includegraphics[width=\columnwidth]{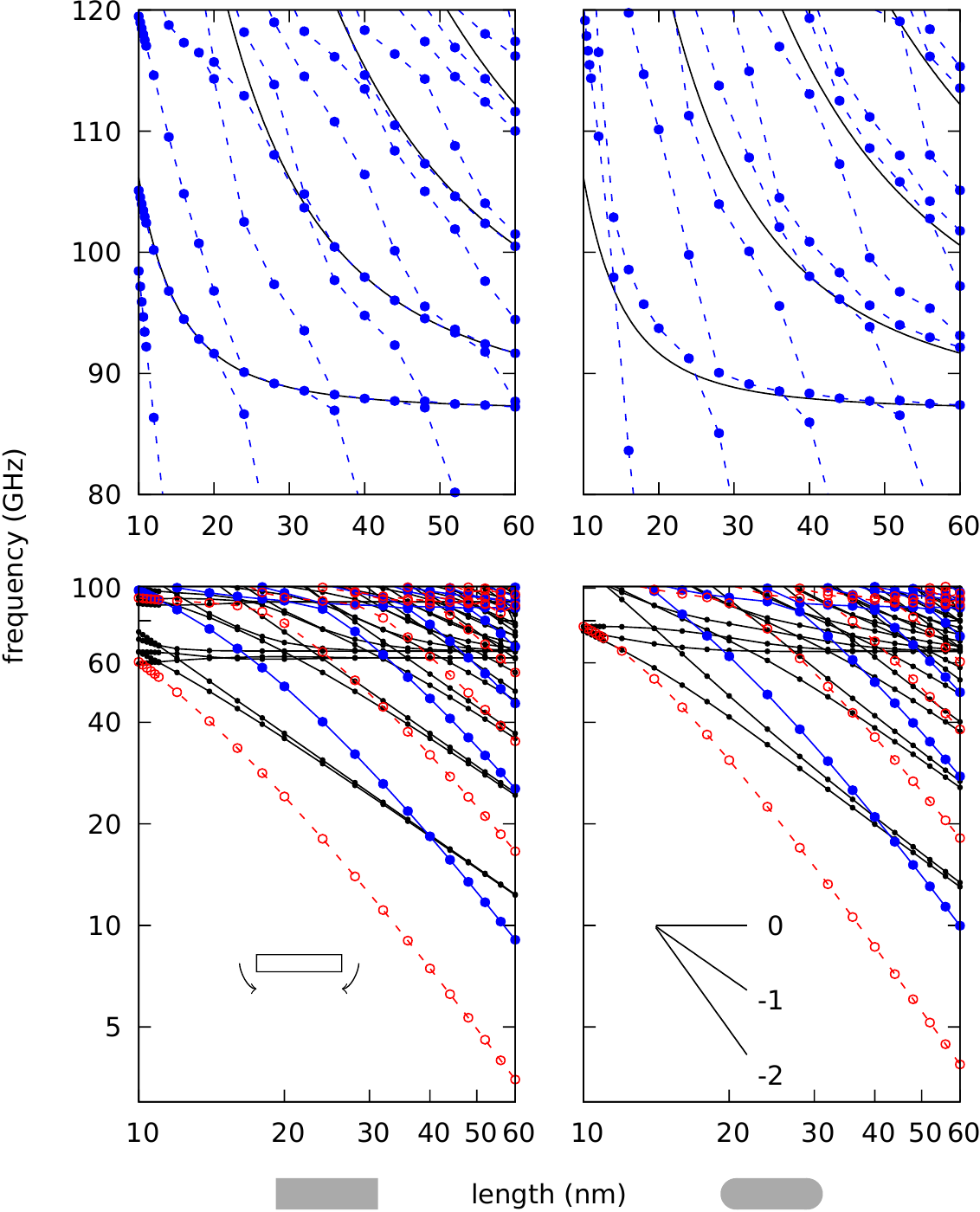}
\caption{Bottom: lowest frequency vibrations of anisotropic circular NRs
as a function of their length $L$.
The NRs ends are flat (left) or half-spheres (right).
The E\textsubscript{g} branches are plotted with full circles and a continuous line (blue online) and
the E\textsubscript{u} branches with empty circles and a dashed line (red online).
All the other modes are plotted with black circles and a continuous line.
Three slopes 0, -1 and -2 are plotted for comparison.\\
Top: E\textsubscript{g} branches in the frequency range of the $q=0$ E\textsubscript{g} vibration of the NW. The continuous black curves are the confined vibrations
calculated from the NW phonon branch at $q=n\frac{\pi}{L}$ with odd $n$.
}
\label{log}
\end{figure}

The third NW phonon branch with a vanishing frequency at $q=0$ is for
flexural phonons.\cite{NishiguchiJPCM97}
It is related to the bending vibrations of the NR with irreducible representations E\textsubscript{u} and E\textsubscript{g}.
These vibrations can be modeled analytically using the Euler-Bernoulli or Timoshenko beam theory.\cite{LeissaJASA1995}
The resulting frequencies vary as $1/L^2$.
This variation is reproduced at large $L$ by considering confined modes at $q=n\frac{\pi}{L}$
because the slope of the NW flexural phonon branch vanishes at $q=0$.
However, the frequencies are significantly different (not shown).
This is because this trivial approach does not reproduce
the displacements obtained in the framework of the Timoshenko beam theory
which contain hyperbolic functions of spatial coordinates ($\sinh$ and $\cosh$).
Therefore, a given NR bending vibration can not be easily matched to a NW flexural phonon
with $e^{iqz}$ dependence and real $q$ .
To check the $1/L^2$ variation, all the lowest frequency branches of the NRs
are plotted in Fig.~\ref{log} (bottom).
Logarithmic scales are used in order to distinguish between different behaviors at large $L$.
Branches whose frequency does not depend on $L$ have a constant frequency (slope 0).
This is the case for phonons confined along a direction perpendicular to the axis of the NRs.
Some of them will be discussed later.
Modes whose frequency vary as $1/L$ have slope -1.
This is the case for vibrations confined along the length of the NRs.
These branches are the extensional and torsional vibrations discussed above.
The remaining branches whose slopes tend to -2 at large $L$
are the E\textsubscript{u} and E\textsubscript{g} bending vibrations.

\subsubsection{Quadrupolar-like modes}
\begin{figure}
\includegraphics[width=\columnwidth]{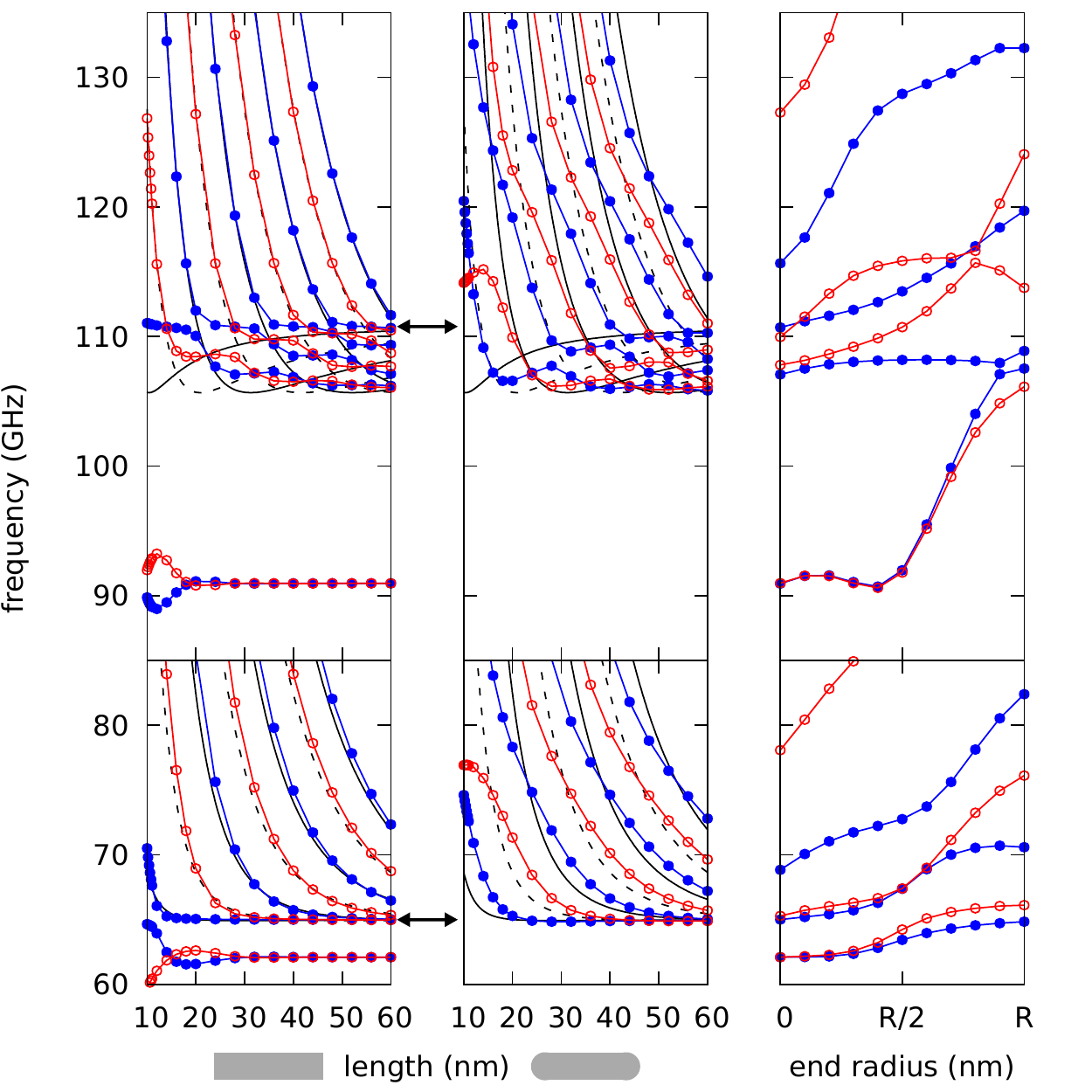}
\caption{Frequencies of the quadrupolar-like vibrations of NRs as a function of $L$.
The left plot is for NRs with flat ends and the center plot for half-sphere ends.
The full circles and lines are for the even (B\textsubscript{1g} (bottom)
and B\textsubscript{2g} (top)) branches (blue online).
The empty circles and lines are for the odd (B\textsubscript{2u} (bottom)
and B\textsubscript{1u} (top)) branches (red online).
The continuous black curves in the left and center plots are the confined vibrations
calculated from the NW phonon branch at $q=n\frac{\pi}{L}$ with odd $n$.
The dashed black curves are for even $n$.
Double-head arrows mark the frequencies of the corresponding quadrupolar-like $q=0$
NW phonons.
The right plot is for $L=\SI{30}{nm}$ with different shapes for the NR ends (see text).
}
\label{quad}
\end{figure}

Let us now consider vibrations of the NRs whose frequency does not tend to zero as $L$ increases and
how they relate or not to phonon branches of the NWs whose frequency does not vanish at $q=0$.
Vibrations similar to the quadrupolar vibrations of the NSs are of course of interest since
they correspond to the most intense low-frequency Raman peak for NSs.
As presented in Table~\ref{brqu}, the branches of interest for the NRs are the B\textsubscript{1g}
and B\textsubscript{2g} ones.
They are plotted in Fig.~\ref{quad}.
The B\textsubscript{1g} vibrations correspond to out of phase elongations along both
\hkl<100> directions perpendicular to the symmetry axis of the NRs.
For the B\textsubscript{2g} vibrations, the elongations are along the \hkl<110> directions.
See Fig.~\ref{disp} (bottom).

The first notable feature is the existence of isolated low-frequency B\textsubscript{1g}
and B\textsubscript{2g} modes for NRs with flat ends (Fig.~\ref{quad} (left)).
Their frequencies at about 62 and \SI{91}{GHz} hardly depend on $L$. 
This is at odd with most other branches which tend to stack up
to form the phonon branches of the NWs as $L \rightarrow \infty$.
These two isolated branches of the NRs can not be related to the NW phonon branches
plotted in Fig.~\ref{disp}.
Looking at the corresponding displacements reveals that the vibrations are localized at both ends of the NRs.
\citet{LibovPMA2013} reported the existence of such localized
vibrations as $m=2$ vibrations for isotropic NRs with flat ends.
In the present case they split due to cubic anisotropy into B\textsubscript{1g} and
B\textsubscript{2g}.

No such isolated branch is observed for NRs with half-spheres at both ends (Fig.~\ref{quad} (center)).
Obviously, the frequency of vibrations localized at the ends depend on the shape of the ends.
Figure \ref{quad} (right) presents the variation of the frequencies of the same vibrations
for $L=\SI{30}{nm}$ as a function of the shape of both ends of the NRs.
The shape is defined by a spheroid with two radii equal to $R$ and the other
radius $R_e$ varying from 0 for flat ends to $R$ for half-sphere ends.
The frequency of the localized vibrations increases with $R_e$
and the isolated branches couple and merge with the other branches.
These localized vibrations can be related to vibrations of the NSs.
By considering the associated displacements and the symmetry of the vibrations,
they were shown to be equivalent to the torsional vibrations
of an isotropic NS with $\ell=3$ and $m=\pm2$
whose frequency is \SI{154.39}{GHz} which is larger than that of the $l=2$ spheroidal mode at \SI{106.20}{GHz}.
They also split into B\textsubscript{1g} and B\textsubscript{2g} when elastic and shape
anisotropies are taken into account.
In all of these cases, the lowest frequency quadrupolar-like vibrations (see Table~\ref{brqu})
have a frequency lower than that of the torsional-like vibrations coming from the $\ell=3$ and $m=\pm2$ vibrations.
Therefore, for NRs with half-sphere ends, there are no separate localized vibrations branches
but rather coupled localized and quadrupolar-like vibrations.

The frequency of confined quadrupolar-like B\textsubscript{1g} and B\textsubscript{2g} vibrations
was determined from the corresponding NW phonon branch using $q=(2 n + 1) \frac{\pi}{L}$.
The corresponding branches are plotted in Fig.~\ref{quad} (left and center).
For NRs with flat ends, the agreement between the frequencies for the NRs
and the ones determined from the NW phonon branches is good for the B\textsubscript{1g} vibrations.
The agreement is excellent for the B\textsubscript{2g} vibrations above \SI{110}{GHz}.
The anti-symmetric vibrations of the same origin (B\textsubscript{2u} for B\textsubscript{1g} and B\textsubscript{1u} for B\textsubscript{2g}) are also plotted for completeness.
The agreement is not so good for NRs with half-sphere ends.
As discussed above, this is due to the coupling with the localized vibrations.
This coupling manifests as complex anti-crossing patterns due to the fact that branches
having the same irreducible representation do not cross.
The same rule explains the complex frequency variations for B\textsubscript{2g} vibrations
between 105 and \SI{110}{GHz}.
Contrarily to the extensional and torsional vibrations discussed before,
the frequency variations of the NW quadrupolar-like B\textsubscript{1g} and B\textsubscript{2g}
phonon branches with $q$ are not monotone.
Their frequencies decrease slightly at small $q$ and then increase.
The decrease is more pronounced for B\textsubscript{2g}.
As a result, the branches for the different overtones of the calculated confined vibrations
cross.
This is hardly noticeable for B\textsubscript{1g} but it is clearly visible for
B\textsubscript{2g}.
This results in anti-crossing patterns which render the assignment of the B\textsubscript{2g} vibrations
to a specific overtone sometimes difficult in this frequency range.
This also explains why the lowest B\textsubscript{2g} frequency does not
match with the frequency of the NW phonon branch at $q=0$.
It corresponds to the lowest frequency of the phonon branch which is
at $\simeq\SI{105.7}{GHz}$, \textit{i.e.}, about \SI{5}{GHz} less than the $q=0$ value.
For B\textsubscript{1g}, this difference is two orders of magnitude less ($\SI{-0.07}{GHz}$).

The quadrupolar vibration of the isotropic sphere (spheroidal modes with $\ell=2$ and degeneracy 5)
split into A\textsubscript{1g}+B\textsubscript{1g}+B\textsubscript{2g}+E\textsubscript{g}
in D\textsubscript{4h}.
We have considered the A\textsubscript{1g} (extensional) and B\textsubscript{1g} and B\textsubscript{2g}
modes above.
The frequencies of the E\textsubscript{g} modes are plotted in Fig.~\ref{log}.
At low-frequency, they correspond to overtones of the bending modes as already discussed.
Anti-crossing patterns appear at \SI{86.9}{GHz} which is the frequency of the
E\textsubscript{g} modes (degeneracy 2, $q=0$ ) of the corresponding NW.
These are thickness-shear vibrations\cite{Eringen} whose displacement is along the $z$ axis of the NW.
One mode corresponds to the $x>0$ part of the NW moving along $+z$ while the $x<0$ part
moves along $-z$. The other mode is obtained by replacing $x$ by $y$.
During such vibrations, the external shape of the NW remains unchanged.
Therefore, a very small coupling with the surface plasmon resonance is expected for the NW.
For NRs, the frequencies of the corresponding modes are plotted in Fig.~\ref{log} (top)
using the same procedure as before.
In that case, the shape of the NRs varies during the vibrations because they correspond
to phonons propagating along $z$ with a non-vanishing wavevector ($q=n\frac{\pi}{L}$)
and also because of the shape of the ends.
Still, the intensity of the corresponding Raman peaks must tend to zero as the aspect ratio of the NRs increases to turn into a NW.

\subsubsection{Breathing-like modes}

\begin{figure}
\includegraphics[width=\columnwidth]{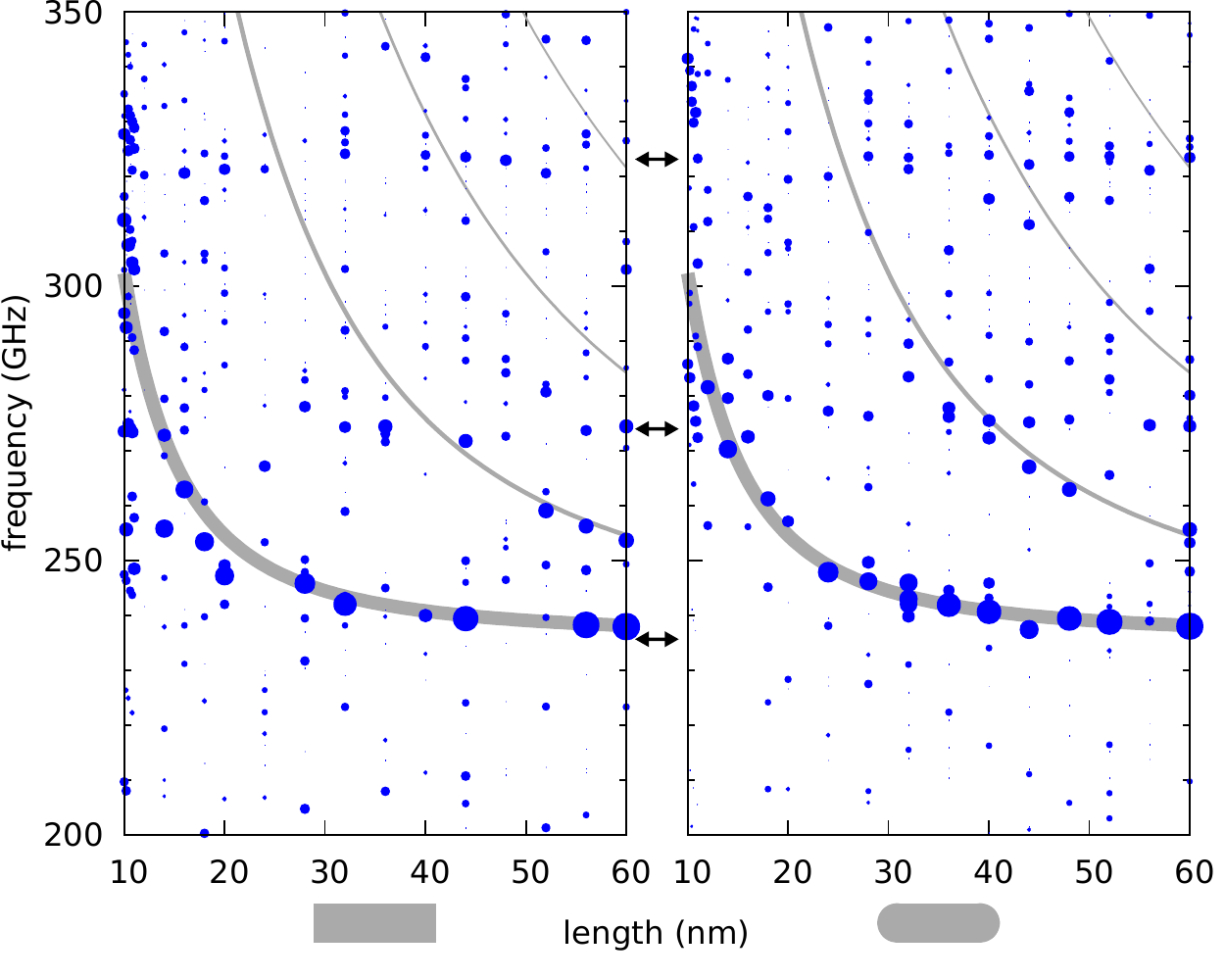}
\caption{Frequency of the A\textsubscript{1g} vibrations of NRs as a function of $L$ compared with the frequency of the confined vibrations calculated at
$q=(2 n + 1)\frac{\pi}{L}$ from the NW breathing-like branch.
The left plot is for flat ends and the right one for half-sphere ends.
The surface area of each circle and the thickness of the curves
are proportional to the variation of the volume during the vibration.
The frequencies of the $q=0$ A\textsubscript{1g} phonons of the NW are indicated with arrows.
}
\label{breath}
\end{figure}

Applying the same procedure to breathing-like vibrations is more challenging because of
the convergence issues and also because of the mixing with other vibrations.
Convergence is very good for the lowest frequency vibrations but it decreases as the vibration index increases.
For $L=\SI{10}{nm}$, the vibration index for the breathing-like vibrations is about 100.
For $L=\SI{60}{nm}$, it is about 350.
This is due to vibrations whose frequency decreases as $L \rightarrow \infty$ to form
the NW phonon branches at lower frequency.
Convergence issues render calculations for large $L$ less reliable.
In addition, several totally symmetric A\textsubscript{1g} branches couple
with the breathing-like vibrations.
They come from the spheroidal vibrations of the isotropic NS with even $\ell$ and $m=0$.
This results in several anti-crossing patterns in the $L$ range investigated in this work.
As seen before, this renders the interpretation in terms of confined vibrations less reliable.
Two modifications are introduced to overcome this issue.
First, the breathing-like vibrations are identified by looking at the volume variations.
This enables to weight the numerous A\textsubscript{1g} vibrations according to their
breathing-like character.
The modes corresponding to the largest circles are those who are expected to add up
to form the feature associated to ``the'' breathing mode in pump-probe or Raman experiments.
In addition, the NW phonon branch used for the calculation of the confined vibrations is
the one which was used as a guide for the eye in Fig.~\ref{disp}.
This results in a simpler picture which is free of anti-crossing patterns.
Fig.~\ref{breath} displays the resulting branches and the confined
vibrations calculated with $q=(2 n + 1)\frac{\pi}{L}$.
As expected, the calculated confined vibrations fail to capture all the details of the complex
coupling between the numerous A\textsubscript{1g} branches.
But the first confined branch matches quantitatively the vibrations having the largest volume variation
for both kinds of NRs.
The following branch with the largest volume variation is also reproduced for $L>\SI{40}{nm}$.
A closer investigation of the anti-crossing patterns in Fig.~\ref{disp} reveals that
the breathing branch couples with two almost flat branches starting from the
next two $q=0$ A\textsubscript{1g} phonons at 274 and \SI{323}{GHz}.
The A\textsubscript{1g} vibrations of the NRs at about these two frequencies
have larger volume variations.
As a conclusion, while a main breathing-like feature definitively exists in Fig.~\ref{breath}
which is correctly described in terms of a fundamental confined vibration,
other factors come into play which may manifest in experiments as close peaks
or a broader peak.

\subsection{Raman intensities}

\begin{figure}
\includegraphics[width=\columnwidth]{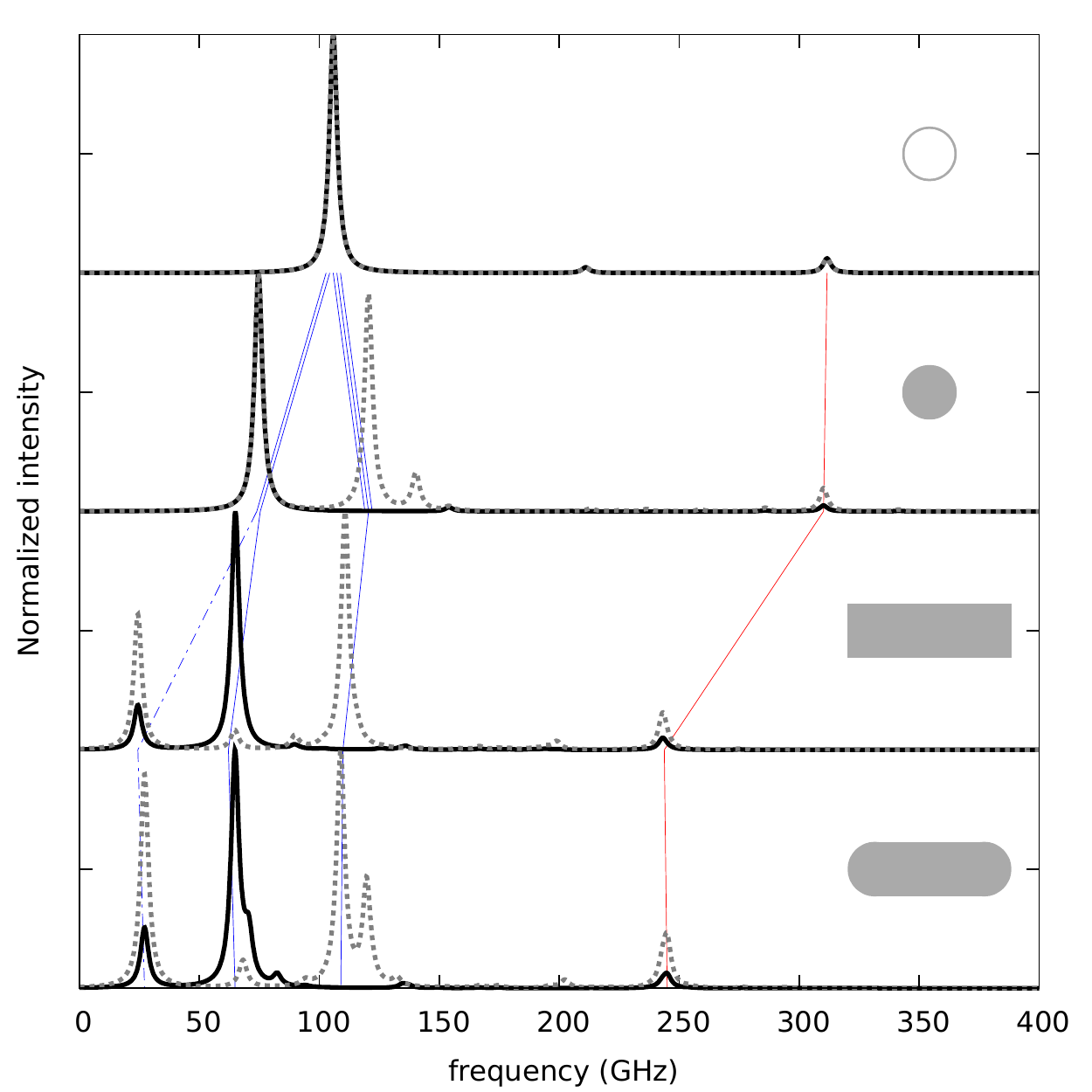}
\caption{Calculated low-frequency Raman spectra for a sphere made of isotropic gold (top)
and a sphere, a cylinder with flat ends and a cylinder with half-sphere ends made
of cubic gold from top to bottom.
The normalized spectra are calculated for the backscattering geometry along \hkl[100] (line)
and \hkl[110] (dashed line) with the light polarization along \hkl[001].
The radius of the spheres and cylinders is \SI{5}{nm}.
The length of the cylinders is \SI{30}{nm} ($L/d=3$).
The lines show the frequency changes and splittings for the quadrupolar vibration (left, blue online)
and the breathing vibration (right, red online) as the elasticity and shape change.
}
\label{intensity}
\end{figure}

In order to illustrate the previous results,
low-frequency Raman spectra have been calculated and are presented in Fig.~\ref{intensity}.
Similar calculations have been carried out until now only for isotropic
spherical NPs\cite{BachelierPRB04,GirardJCP17} using analytic expressions for the
vibrations (Lamb modes) and the electric field inside the NPs (Mie solutions).
Analytic expressions are not available for the anisotropic and
non-spherical NPs considered in this work.
To overcome this problem, the spectra in Fig.~\ref{intensity} were calculated according
to the method described in Ref.~\onlinecite{GirardJCP17} (equations 11 and 12),
using the vibrations obtained previously and assuming a constant electric field inside the NPs.
As a result, the calculation of the intensity comes down to evaluating the Brillouin scattering term
of Ref.~\onlinecite{MontagnaPRB08}, \textit{i.e.}, a volume integral involving only the displacement
field and the retardation effect ($e^{-i\mathbf{q.R}}$).
By expanding this last term as $1-i\mathbf{q.R}+\cdots$, the calculation only requires
to evaluate volume integrals of $x^l y^m z^n$ functions for which analytical expressions
exist as in the Rayleigh-Ritz approach presented before.

The validity of this approach is dubious since the variation of the electric field inside the NPs is not taken into account.
For example, it fails to reproduce the depolarized scattering by the quadrupolar-like modes.
Still, the spectrum calculated for the isotropic spherical NP is similar to the one
in Ref.~\onlinecite{GirardJCP17}.
For this reason, the calculated spectra are expected to provide an insightful ``first-order''
approximation of actual Raman measurements.

The spectra in Fig.~\ref{intensity} were calculated using all the vibration modes
confirming that only the Raman-active ones contribute to the spectra.
All the Raman peaks were convoluted by a Lorentzian function (full width at
half-maximum \SI{2}{GHz}).
The calculations were performed for the backscattering geometry along \hkl[100] (line)
and \hkl[110] (dotted line) with the light polarization along \hkl[001].
They clearly confirm the previously discussed features, namely the presence of intense
peaks for the quadrupolar-like A\textsubscript{1g} (extensional), B\textsubscript{1g}
and B\textsubscript{2g} modes, no significant scattering from the localized vibrations (flat ends)
and the E\textsubscript{g} modes, and small peaks for the breathing-like vibrations.
Deviations from these calculations are expected because of the variations of the
electric field inside the nanoparticles even for isotropic spheres.\cite{GirardJCP17}
Larger deviations can also occur when NPs are close enough that the surface plasmons
of neighbor NPs couple. In that case, new low-frequency Raman peaks may appear as
recently observed in gold nanoparticles super-molecules.\cite{GirardNL16}

\section{Conclusion}

The vibrations of gold NRs have been investigated in order to point out
the experimentally relevant ones and in particular those which are
expected to have the largest Raman cross sections.
To this end, gold NRs and NWs have been considered
using the Rayleigh-Ritz variational method.
The symmetry and the volume variation of the modes were determined.
Raman spectra have also been calculated by considering only the Brillouin scattering mechanism.
Elastic anisotropy is shown to play a major role for NWs made of single domain cubic gold. 
The vibrations of NRs with flat ends have been compared with confined vibrations obtained as
phonons of the NW at fixed wavevector $q=n\frac{\pi}{L}$.
The frequencies of most vibrations of the NRs can be estimated quickly and rather accurately using
only the dispersion curves of the NW and the length of the NRs.
This simple picture enabled to understand qualitatively and often quantitatively the origin
of the main Raman active vibrations including the totally symmetric ones which are of
interest in time-domain measurements.
Localized B\textsubscript{1g} and B\textsubscript{2g} vibrations have been identified.
Their Raman scattering cross-sections are small because they are related to torsional vibrations of the NSs.
However, the Raman intensities calculated in this work would have to be reconsidered
in particular when surface enhanced Raman scattering conditions are met,
for example when the ends of neighbor NRs are very close.
This study also quantifies the influence of the shape of the ends of the NRs.
Differences between flat and half-sphere ends are small as far as fundamental vibrations are concerned.
Larger differences exist for overtones.
Vibrations localized at the ends of the NRs are also strongly affected.
However, observing these differences in experimental measurements is very challenging.
On the opposite, the major role played by elastic anisotropy has been highlighted.
The resulting splitting of the quadrupolar-like vibrations is expected to be a clear signature
of the single domain cubic gold inner structure as already reported for NSs.\cite{PortalesPNAS08}

\bibliography{cyl}

\end{document}